\documentclass{aipproc}

\layoutstyle{8x11single}
\SetInternalRegister\hbadness{8000} % pseudo latin isn't breaking very well :-)

\newcommand\doingARLO[2][]{%
  \ifx\mmref\undefined #1\else #2\fi
}

\begin{document}

\title
      [Workshop summary and the future]
      {Workshop summary and the future}


\author{Gianfranco De Zotti}{
  address={Osservatorio Astronomico di Padova, Vicolo dell'Osservatorio 5, I-35122 Padova, Italy},
  email={dezotti@pd.astro.it},
  thanks={\ }
}

\iftrue
\author{Carlo Burigana}{
  address={ITeSRE-CNR, Via P. Gobetti 101, I-40129 Bologna, Italy},
  email={burigana@tesre.bo.cnr.it},
}

\author{Carlo Baccigalupi}{
  address={SISSA, Via Beirut 4, I-34014 Trieste, Italy},
  email={bacci@sissa.it},
  homepage={\ },
  altaddress={\ }{\ }
}
\fi

% \copyrightholder{Acoustical Society of America}
\copyrightyear  {2001}

\def\lsim{\, \lower2truept\hbox{${< \atop\hbox{\raise4truept\hbox{$\sim$}}}$}\,}
\def\gsim{\, \lower2truept\hbox{${> \atop\hbox{\raise4truept\hbox{$\sim$}}}$}\,}

\begin{abstract}
%\thanks{A footnote in the abstract}
%\thanks{A second footnote in the abstract}
We present a tentative summary of the many very interesting issues
that have been addresses at this workshop, focussing in particular
on the perspectives for measuring the polarization power spectra
of the Cosmic Microwave Background produced by scalar and tensor
perturbations, in the presence of foregrounds.

\end{abstract}

\date{\today}

\maketitle

\section{Introduction}

%\ifthenelse{\equal\selectedlayoutstyle{6x9}}{\par\bfseries text
%\par\bfseries\par\bfseries\normalfont}{}

% Some url test \url{http://www.world.universe}.
The astonishing advances in our understanding of the basic
properties of the Universe and in precision determinations of its
fundamental parameters, made possible by the recent accurate
measurements of acoustic peaks of the Cosmic Microwave Background
(CMB) anisotropy power spectrum by the TOCO \cite{Miller1999},
Boomerang \cite{deBernardis2000, deBernardis2001}, Maxima
\cite{Hanany2000}, DASI \cite{Halverson2001}, and CBI
\cite{Padin2001} experiments, have strongly highlighted the
extraordinary wealth of cosmological information carried by the
CMB. As the ongoing MAP mission %\url{http://map.nasa.gsfc.gov}
and the forthcoming {\sc Planck} satellite
%\url{http://astro.estec.esa.nl/Planck} will pin down some 10 key
%cosmological parameters with exquisite accuracy and
will provide high sensitivity, high resolution all sky CMB
temperature maps
%us with a detailed insight on the origin and evolution of structure
%in the Universe,
the new frontier has become CMB polarization, first investigated
by \cite{Rees1968} and \cite{Kaiser1983}.

The information content of the CMB polarization field is in fact
richer than that of the temperature field since, in addition to
amplitude it also has an orientation. Recent analyzes
\cite{Seljak1997, Kamionkowski1997, ZaldarriagaSeljak1997} have
shown that, rather than with the conventional Stokes parameters Q
and U (the Stokes parameter V describing circular polarization can
generally be neglected because it cannot be generated through
Thomson scattering), it is more convenient to work with the two
rotationally invariant fields $E$ and $B$, which are linear, but
non-local, combinations of Q and U. This decomposition of the
$2\times 2$ symmetric trace-free tensor describing the linear
polarization state is analogous to that of a vector field into a
curl and a curl-free (gradient) component (e.g.
\cite{KamionkowskiJaffe2000}).

If the CMB fluctuations are Gaussian, they can be fully
characterized by four power spectra, $C_\ell^{TT}$, $C_\ell^{EE}$,
$C_\ell^{BB}$, $C_\ell^{TE}$. Under reflections (parity
transformations) $E$ transforms as a scalar (like $T$), while $B$
transforms as a pseudo-scalar; therefore the cross-correlations
$TB$ and $EB$ vanish, as discussed by {\it Sahzin}. The only fact
that the temperature field yields just the first of those
power spectra indicates how CMB polarimetry increases the
cosmological information. But there is much more than that: scalar
perturbations, held responsible for the acoustic peaks in the
temperature power spectrum, do not produce, to first order,
$B$-mode signal. The $B$-mode component may contain the signature
of tensor perturbations (gravity waves) produced during the
inflationary epoch and allow to measure the inflaton potential
height\footnote{A $B$ component can also be generated by vector
perturbations that may be excited by topological-defect models but
are not present in inflationary models} (see, e.g.
\cite{KamionkowskiKosowsky1999})! Its detection would therefore
give us a glimpse of the Universe at $10^{-38}$ seconds after the
initial singularity, at energy scales of $\sim 10^{16}\,$GeV, many
orders of magnitude above those accessible at accelerators, and
would therefore have profound implications not only for cosmology
but also for particle physics \cite{Peterson1999}.

Measurements of CMB polarization are however extremely challenging
because the signal is very weak (at several $\mu$K level on small
angular scales and much less on large angular scales) and liable
to be strongly contaminated by polarized foreground emissions. The
definition of suitable strategies therefore requires a close
collaboration between experiment builders, theorists, experts on
the various relevant Galactic and extra-galactic foregrounds. This
very timely workshop has offered a badly needed occasion for
experts in the different fields to meet.

\begin{figure}
\resizebox{\hsize}{15pc}
%\raisebox{0truecm}[0truecm][-2truecm]
%{\resizebox{0.5\textheight}{!}
%\psfig{file=cmb_pol.ps,height=4.in}
%{\resizebox{\hsize}{!}
%\resizebox{\hsize}{!}
{\includegraphics{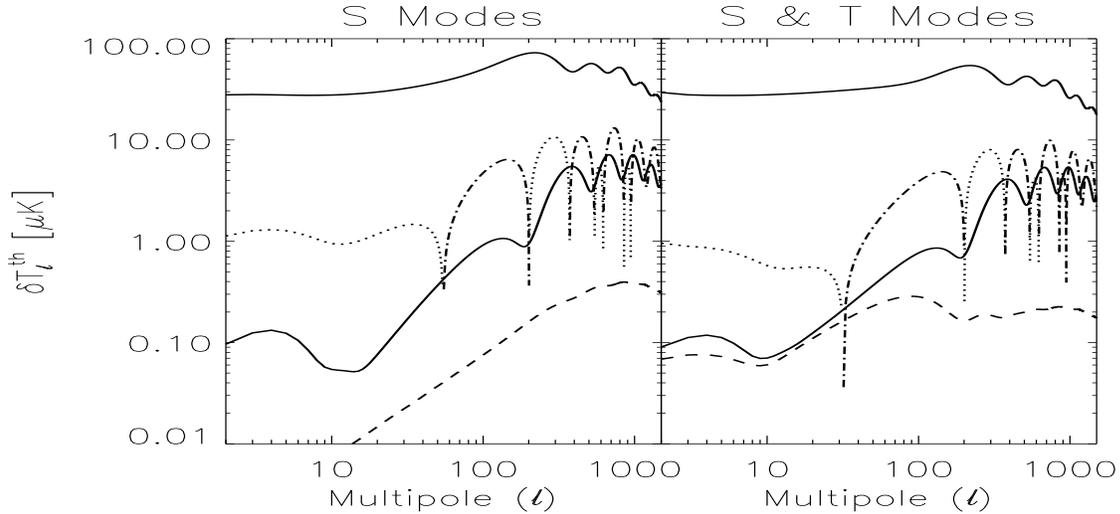}}%\vspace{-5truecm}}
%\includegraphics[height=.5\textheight]{cmb_pol.ps}
%\raisebox{-2truecm}[0truecm][-2truecm]
{\caption{Temperature (upper solid line) and polarization power
spectra for the cosmological model specified in the text. The
panel on the left corresponds to pure scalar perturbations, the
one on the right to scalar plus tensor perturbations yielding
equal contributions to the temperature quadrupole. The partly
dotted, partly dot-dashed lines represents the $TE$
cross-correlation; the dot-dashed portion shows the absolute value
of the cross-correlation when it is negative. The lower solid
lines and the dashed lines represent $E$ and $B$ power spectra,
respectively. In the case of scalar perturbations, a $B$
contribution is originated by gravitational lensing. In the case
of scalar plus tensor perturbations, gravitational lensing adds a
contribution showing up at large $\ell$.}}
\end{figure}

\section{CMB polarization power spectra }

%\subsection{Small angular scales}

CMB polarization is induced by Thomson scattering of anisotropic
radiation with a quadrupole pattern in the rest frame of the
electron (see e.g. \cite{HuWhite1997}). Before recombination,
anisotropies were strongly damped out by the tight coupling
between photons and baryons, so that the polarization that could
be generated was correspondingly depressed. To generate a
quadrupole, a gradient in the velocity of the photon-baryon
fluid across the photon mean free path, $\lambda_p$, is
necessary: only perturbations on scales small enough to produce
anisotropies on scales $< \lambda_p$ can give rise to
polarization. %But on small scales perturbations are damped by
%photon di{f}fusion.
As recombination proceeds, $\lambda_p$
increases rapidly and polarization is produced. The polarization
degree increases with decreasing angular scale (increasing $\ell$)
reaching a maximum of $\sim 10\%$ on the minimum scales that
survived damping prior to recombination \cite{HuDodelson2001}.

Figure 1 shows simulated temperature, polarization and $TE$
correlation power spectra (discussed by {\it Balbi}), in terms of
thermodynamic temperature fluctuations per logarithmic interval
in $\ell$, $\delta T =
\left[\ell(2\ell+1)C_\ell/4\pi\right]^{1/2}$, for a flat
$\Lambda$CDM cosmology with $\Omega_{\rm baryon} = 0.03$,
$\Omega_{\rm CDM} = 0.3$, $\Omega_{\Lambda} = 0.67$,
$H_0=65\,\hbox{km}\,\hbox{s}^{-1}\,\hbox{Mpc}^{-1}$, 3 species of
massless neutrinos, $Y_{\rm He} = 0.24$ and a re-ionization
optical depth $\tau=0.04$. Calculations have been made using the
CMBFAST code by \cite{SeljakZaldarriaga1996}. The panels show the
cases of purely scalar, scale invariant, isentropic primordial
perturbations (left), and of scalar plus tensor perturbations with
a unit ratio of tensor ($T$) to scalar ($S$) contributions to the
temperature quadrupole (right). A $T/S \simeq 1$ is probably an
upper limit, once all the relevant constraints are taken into
account \cite{Zibin1999}.

As illustrated by Fig.~1, the ($E$-mode) polarization
fluctuations associated to acoustic peaks have relatively more
power on small scales than temperature fluctuations. Also, since
polarization is related to peculiar velocities on the last
scattering surface, the peaks of its power spectrum are out of
phase by $\pi/2$ with those of temperature anisotropies
(velocities are out of phase with density perturbations, like
velocity and position of a harmonic oscillator). The $TE$ power
spectrum, being the product of the two, oscillates at twice the
frequency. The relative position of the temperature and
polarization peaks is thus the signature of coherent perturbations
(as opposed to causal mechanisms, e.g. those due to topological
defects), hence of inflation, as the origin of structure. If
large scale structure was produced by some causal mechanism, the
first peak would have to occur at smaller angular scales in
order to be within the causal horizon at last scattering
\cite{SpergelZaldarriaga1997}. %Therefore, just the detection of
%the peaks of  the strongest CMB polarization component, induced by
%scalar perturbations, will allow an unambiguous test on the nature
%of primordial perturbations.

A lot more can be learned from an accurate determination of the
power spectrum of the $E$-mode on small angular scales, and, to
some extent, to the $T$-$E$ correlation power spectrum
\cite{Zaldarriagaetal1997, Kinney1998, Eisensteinetal1999,
Enqvist2000, Bucher2001}. For example, polarization measurements:
identify contributions from isocurvature perturbations that can be
confused with tensor perturbations or early re-ionization effects
on temperature data; constrain scalar-tensor theories of gravity
which may generate scalar, vector and tensor modes, leaving
distinctive signatures in the CMB; are sensitive to effects of
quintessence models; increase significantly the precision of the
determination of most cosmological parameters; are directly
informative on the details of the recombination and, therefore, on
processes governing it \cite{Naselsky2001}.

%\subsection{Large angular scales}
As mentioned above, the amplitude of the polarization power
spectrum on large angular scales originated at recombination is
expected to be very small because the photon mean free path is
small and multiple scattering erase polarization. On the other
hand, re-ionization strongly increase the the polarization signal
at low $\ell$ \cite{Zaldarriaga1997,KamionkowskiKosowsky1998}
producing a characteristic bump (see Fig. 1) at $\ell \sim
\sqrt{z_{\rm ri}}$, $z_{\rm ri}$ being the redshift of
re-ionization,  with amplitude roughly proportional to the optical
depth for Thomson scattering.
%
%\begin{equation}
%\tau = 0.041{h\Omega_b\over \Omega_m}\left\{\left[1-\Omega_m
%+\Omega_m(1+z_{\rm ri})^3\right]^{1/2} -1 \right\}\ .
%\end{equation}
%
The polarization power spectrum at low $\ell$ is therefore a
sensitive probe of re-ionization.

The lack of an observed Ly$\alpha$ (Gunn-Peterson) trough in the
spectrum of the SDSS quasar at $z=5.8$ \cite{Fan2000} implies, for
the above choice of cosmological parameters, $\tau > 0.022$.
Re-ionization, however, suppresses temperature fluctuations on
small scales: the lack of observed suppression of the acoustic
peaks gives $\tau \lsim 0.2$ \cite{Wang2001}, implying $z_{\rm ri}
\lsim 20$. Very recently \cite{Becker2001} reported compelling
evidence of a Gunn-Peterson trough in a $z=6.28$ quasar discovered
by the Sloan survey (SDSS), suggesting that $z_{\rm ri} \simeq 6$
(see also \cite{Fan2001b}), implying $\tau$ close to the lower
limit. Keck spectroscopy by \cite{Djorgovski2001} have shown
evidence of dark portions of the spectrum of an SDSS quasar at
$z=5.73$, indicating a dramatic increase of the Ly$\alpha$ opacity
at $z\gsim 5.2$, as expected if we are approaching (from lower
redshifts) the re-ionization era of a non-uniform medium, with
ionized regions intermingled with islands of neutral gas. Clearly,
it may be premature to draw very definite conclusions. Only a
tiny fraction ($\sim 2\times 10^{-5}$) of baryons need to be in
the form of neutral hydrogen to account for the Gunn-Peterson
optical depth implied by the %\cite{Becker2001}
data, and the re-ionization may be non-uniform, so that the
line-of-sight where the e{f}fect was observed may be not
representative.

On the other hand, given the rapid evolution of the optical depth
that is observed, it seems unlikely that the e{f}fective
re-ionization redshift is $\gg 6$. If so, the re-ionization bump
shown in Fig.~1, which corresponds to $\tau=0.04$, is probably an
upper limit. CMB polarization fluctuations on large angular
scales are thus probably at the $0.1\,\mu$K level and their
detection is therefore very hard. Since the SDSS is expected to
find $\sim 20$ quasars at $z\gsim 6$ \cite{Fan2001a}, more
detailed insight into the re-ionization process is expected in the
next few years.

The CMB polarization power spectrum provides additional
information compared to observations of the Gunn-Peterson trough.
While the latter measures just $z_{\rm ri}$, the former also
measures $\tau$, in addition to allowing an estimate of $z_{\rm
ri}$ (from the position of the peak), thus providing a
determination of the baryon fraction in the intergalactic medium,
hence, e.g., of the galaxy formation efficiency, a piece of
information very difficult to derive with other means.

%\subsection{$B$-mode power spectrum}

The specific signature of tensor and vector perturbations is
$B$-type polarization, which cannot be produced by scalar
perturbations because of their symmetry properties. This opens the
exciting possibility of a direct investigation of tensor
components, in spite of the fact that their contribution to the
CMB polarization is expected to be much smaller than that of
scalar perturbations. Since the amplitude of the $B$-mode
component due to gravity waves is proportional to the square of
the inflationary energy scale \cite{ZaldarriagaSeljak1997,
Kamionkowski1997}, its detection amounts to a measurement of the
energy scale of inflation!

For roughly scale invariant tensor perturbations the $B$-mode
power spectrum peaks at ($\ell \simeq 100$; see Fig.~1). As
pointed out by \cite{ZaldarriagaSeljak1998}, gravitational lensing
by the matter distribution, in addition to smoothing the acoustic
peaks of both temperature and polarization, distorts the
polarization pattern on the sky, mixing $E$ and $B$ modes.
$B$-type polarization is thus generated even in the case of pure
scalar perturbations. The effect peaks on small angular scales
($\ell \sim 1000$) and does not interfere with the possibility of
measuring the $B$-mode power spectrum from gravity waves.

Of course, gravity waves are not the only primary sources of
$B$-mode polarization. Primordial magnetic fields ({\it
Wielebinski}) would also produce tensor and vector metric
perturbations, resulting in further CMB fluctuations
\cite{Mack2001, Kahniashvili2001}. A distinctive feature of such
fields is that, for a range of power spectra of perturbations
produced by them, the polarization fluctuations are comparable
to, or larger than, the corresponding temperature fluctuations.
The amplitude of the CMB power spectra vary as the square of the
energy density, i.e. as the 4th power of the magnetic field
amplitude; the effects on the CMB of primordial stochastic
magnetic fields with comoving amplitude much below $10^{-9}$
Gauss are essentially undetectable. Cosmological magnetic fields
may also be detectable through the (frequency dependent) Faraday
rotation of CMB polarization. Faraday rotation also mixes $E$ and
$B$ modes. These effects are however small for realistic magnetic
field strengths \cite{KosowskyLoeb1996, Kahniashvili2001}.

\begin{figure}
%  \resizebox{20pc}{!}{\includegraphics{pol_fore_3freq.ps}}
\resizebox{\hsize}{15pc}{\includegraphics{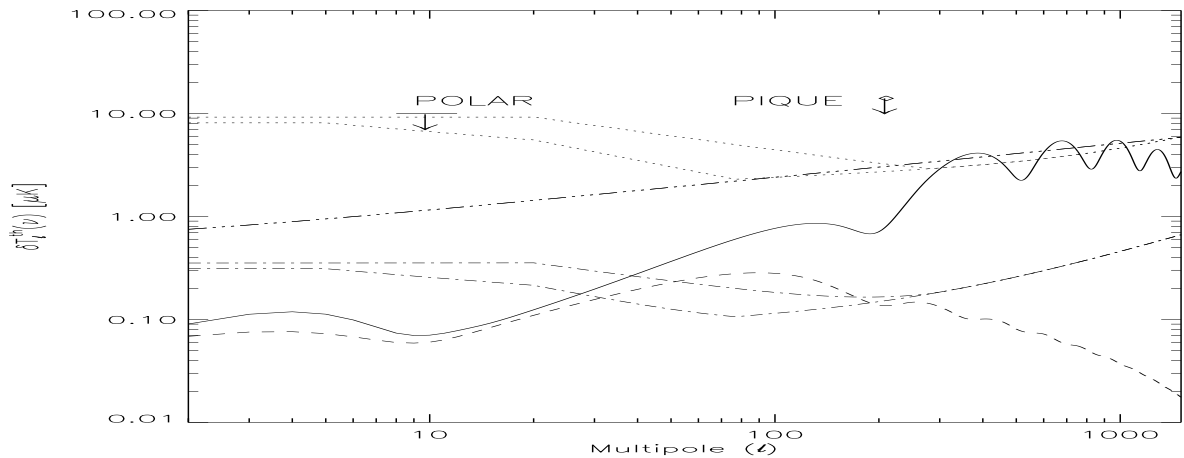}}
\caption{CMB versus foreground polarization power spectra. For CMB
we have plotted the scalar plus tensor model of Fig.~1, but
without gravitational lensing. The foreground power spectra are
shown for 3 frequencies: 30 (dots), 100 (dot-dash), and 217 (three
dots - dash) GHz. They comprise the contributions of Galactic
synchrotron and thermal dust emission, and of extragalactic
sources (radio sources and dusty galaxies); polarization
fluctuations of spinning or magnetized grains are not included.
At 30 and 100 GHz, the dominant polarized foregrounds are
synchrotron for $\ell \lsim 1000$ (the limit actually decreases
with increasing frequency) and extragalactic radio sources at
higher $\ell$. The two lines at these frequencies for $\ell \lsim
200$ bound the range of synchrotron power spectra reported by {\it
Burigana} at this meeting; the synchrotron power spectrum for
higher $\ell$ is from \cite{Baccigalupi2001}. At higher
frequencies, the dominant polarized foreground is expected to be
thermal dust emission, as illustrated by the line at 217 GHz. We
have assumed that the global net polarization degree of dusty
galaxies is 0.4\%, as found by {\it Greaves} (these proceedings)
for M82; if so their contribution to polarization fluctuations
can be significant only on very small angular scales. The $E$
and $B$ mode power spectra for these foreground components are
found to be almost equal; only one of them is plotted. The recent
upper limits set by the POLAR \cite{Keating2001} and the PIQUE
\cite{Hedman2001} experiments are also shown. }
\end{figure}

\begin{figure}
  \resizebox{\hsize}{15pc}{\includegraphics{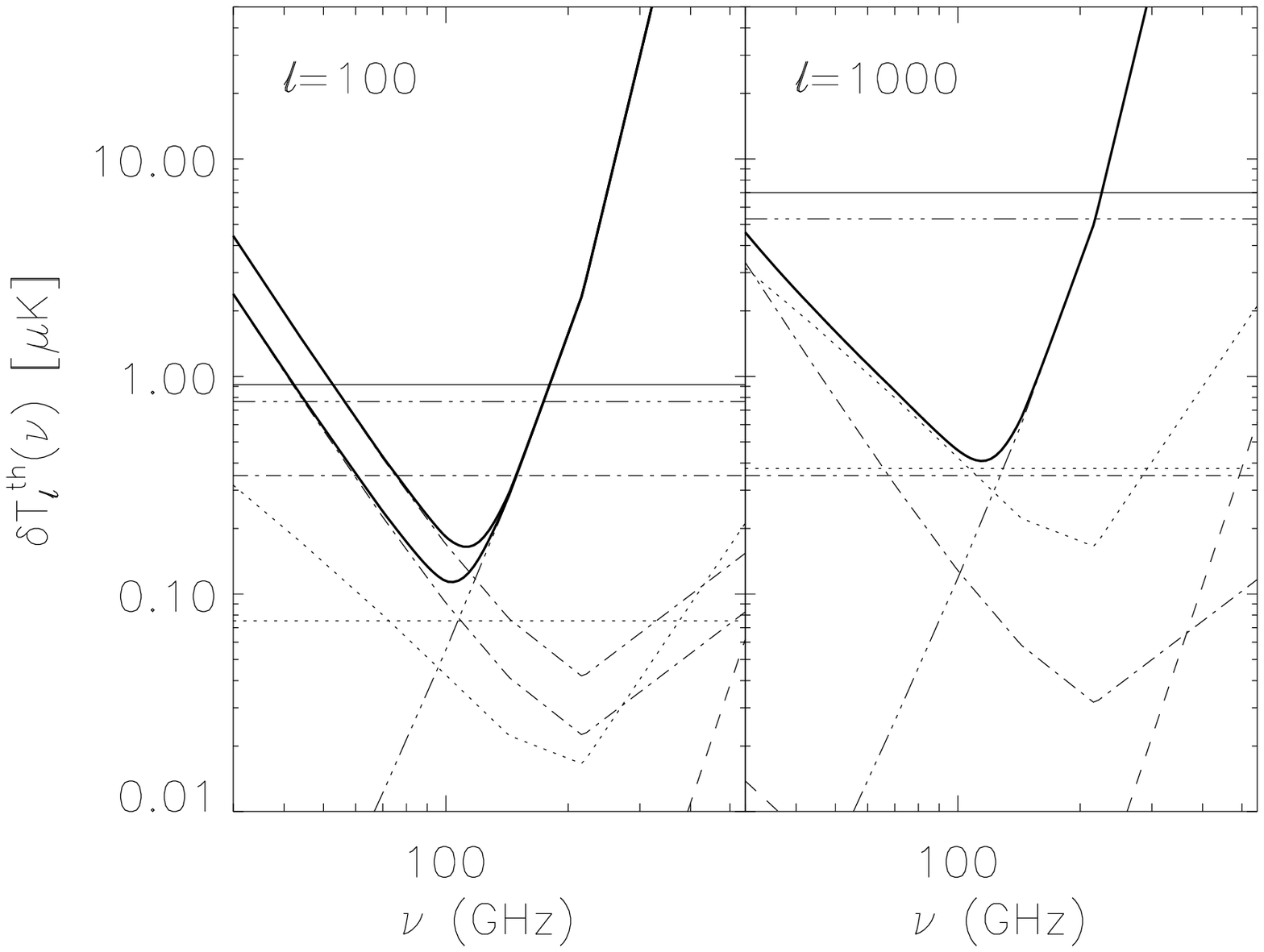}}
\caption{Frequency dependence of foreground polarization
fluctuations for $\ell =100$ (left) and $\ell =1000$ (right).
The dot-dashed curves represent Galactic synchrotron. The two such
curves shown in the left-hand panel bracket the estimates reported
by {\it Burigana} (see caption to Fig.~2). The synchrotron curve
on the right-hand panel is from \cite{Baccigalupi2001}. The
three-dots/dash curve is for thermal dust emission
\cite{Prunet1998}, while the dotted curve is for extragalactic
radiosources \cite{dezotti1999}, which, for $\nu \lsim 100\,$GHz,
dominate foreground fluctuations on small angular scales, as
illustrated by the $\ell=1000$ panel. The dashed line in the
right-hand corner of each panel is for dusty galaxies, assumed to
have a net polarization degree of $0.4\%$. The solid curves show
the sum of the various foreground contributions. For $\ell=100$,
and in general for $\ell \lsim 200$, and $\nu \lsim 100\,$GHz, the
two solid lines reflect the range of estimates for the dominant
synchrotron contributions. The horizontal lines show the CMB
polarization fluctuations for the cases shown in Fig.~1, namely:
$E$-mode for pure scalar fluctuations (solid), $E$-mode for pure
scalar plus tensor fluctuations (three dots/dash), $B$-mode from
the tensor component in the case of scalar plus tensor
fluctuations (dot/dash), $B$-mode due to gravitational lensing
for pure scalar fluctuations (dotted). }
\end{figure}

\section{Polarized foregrounds}

Foregrounds may be a more serious hindrance for CMB polarization
measurements than for temperature measurements because at least
some of them are more polarized than the CMB. At the moment they
are very poorly understood, so that it is still unclear to what
extent they will limit our ability to measure CMB polarization.
Particularly uncertain is the polarized emission from thermal
\cite{Prunet1998}, spinning and ferro- or ferri-magnetic dust
grains \cite{DraineLazarian1999}. %Much more work is needed in this
%area.

Recent analyzes of polarization fluctuations due to the various
foregrounds have been presented and discussed by several speakers
({\it Baccigalupi, Burigana, Fosalba, Lazarian, Ponthieu, Prunet,
Sazhin, Tucci}). Figures 2 and 3 attempt to summarize of the main
conclusions.  At frequencies $\lsim 100\,$GHz foreground
polarization fluctuations are likely dominated by Galactic
synchrotron emission on large angular scales and up to $\ell \sim
1000$ (the limit actually decreases with increasing frequency) and
by extragalactic radio sources on small scales. Note that
synchrotron emission seems to have much more small-scale structure
in polarization than in total intensity, at least in the
relatively low-frequency maps currently available. Such structure
may be due, to some extent, to differential Faraday rotation, and
might therefore not be present in high frequency maps. Spinning
dust grains may be important contributors particularly at
20--$30\,$GHz, while their polarized emission is probably
negligible above $40\,$GHz. Polarized microwave emission from
magnetic grains may be important even if they do not dominate the
overall emission. At higher frequencies, the main polarized
foreground is expected to be, on all scales, thermal dust
emission. In fact, the (still scanty) data on sub-mm polarization
of dusty galaxies ({\it Greaves, Scott}) indicate that their
global net polarization, is likely $< 1\%$. Polarization maps of
the nearest starburst galaxy, M82 \cite{Greaves2000} show values
from 0.7\% to 9.1\%. However, when the contributions of the
differently oriented magnetic vectors are summed together, the
global net polarization is found to be $\simeq 0.4\%$ ({\it
Greaves}).

Figure 2 also suggests that the POLAR experiment
\cite{Keating2001}, at 30 GHz, is already close to detecting
synchrotron polarization fluctuations. There is apparently
little hope of detecting large scale CMB polarization
fluctuations at this frequency, and even at higher frequencies,
a delicate subtraction of the synchrotron emission will be
probably necessary. Such subtraction is further complicated by the
substantial variations of the synchrotron spectral index across
the sky \cite{Platania1998}.

As illustrated by Fig.~3, the foreground contamination is expected
to be minimum at frequencies $\simeq 120\,$GHz. In the optimal
frequency window, foregrounds should not seriously limit our
ability to measure the $E$-mode power spectrum except for large
($\ell \lsim 40$) and very small ($\ell \gsim 2000$) angular
scales.

As pointed out by \cite{Seljak1997}, in most cases foregrounds
yield essentially equal contributions to the $E$ and $B$ power
spectra. %This is obviously the case for uncorrelated point
%sources. In the case of Galactic synchrotron and dust emission,
%the alignment is primarily determined by magnetic fields, which
%are obviously not scalar in nature;
The analysis of low-frequency Galactic polarization surveys and of
extragalactic radio sources, presented by {\it Baccigalupi},
indeed yields essentially equal $E$- and $B$-mode power spectra in
both cases. Since, the CMB polarization is, on small angular
scales, predominantly $E$-mode, the difference between $E$ and $B$
power spectra may be a direct measure of the CMB signal.

The situation is clearly much more difficult for the $B$-mode,
which, even in the optimal frequency range, appears to dominate
foreground fluctuations only if the amplitude of the tensor
component is close to current upper limits, and only over a rather
narrow range of multipoles. Ongoing measurements and the
forthcoming multifrequency maps provided by MAP and {\sc Planck}
will be essential to design future high sensitivity CMB
polarization experiments.

\subsection{Astrophysical information from polarization surveys}

Clearly, the study of the foreground sources mentioned above is of
great astrophysical interest per se, since it will provide crucial
information on their physical properties. Still in the framework
of CMB experiments, as noted above, some polarized astrophysical
sources (Galactic synchrotron, extragalactic radio sources) may be
more easily extracted from polarization than from temperature maps
because their polarization degree may be higher than that of the
CMB. This is particularly true for $B$-mode maps: CMB
fluctuations are very small even in the ``cosmological window''
(50--200 GHz) where they dominate by far temperature maps.
% this may ease quite a lot point source extraction.

%\subsubsection{Galactic polarized emission}

High resolution, high sensitivity polarization surveys of the
Galaxy are becoming available (see contributions by {\it Reich,
Gaensler, Landecker, Sault, Vinyakin}), with particularly detailed
imaging of the Galactic plane. These maps effectively amount to a
tomography of the ISM, to quote {\it Wielebinsky}, and allow to
investigate the ecosystem of the Galaxy, in {\it Landecker}'s
words. In fact these observations are informative on the large
scale structure of the Galactic magnetic field, its structure in
the solar neighborhood in relation with ISM features ({\it
Haverkorn})
%including signatures of energy deposition
(cloud complexes, bubbles, loops, shells, SN remnants, etc.), its
$z$-height structure and the relation between the magnetic field
within the thin ISM disk and the thicker synchrotron emitting
disk, the relation between the magnetic field  within
interstellar clouds and their density and velocity structure, the
role of the magnetic field in regulating the star formation
efficiency, dust properties and dust alignment mechanisms (from
polarization of dust emission). Our current understanding of the
large scale structure of the Galactic magnetic fields has been
reviewed by {\it Han}.

%\subsubsection{Extragalactic sources}

The available information on polarization properties of
extragalactic sources is still very limited particularly at sub-mm
wavelengths ({\it Scott, Greaves}), but not much is available also
for radio sources at $\gsim 10\,$GHz. There is obviously a lot of
astrophysics to be learned from polarization measurements,
particularly at mm/sub-mm wavelengths where the effects of
internal synchrotron self-absorption and of Faraday rotation can
be (with rare exceptions) ignored so that we can reliably assume
that the magnetic field direction lies perpendicular to the
observed polarization position angle.

Another tiny signal with a large information content is the
polarized component of the Sunyaev-Zeldovich effect
\cite{SunyaevZeldovich1980}. It will be important to assess the
effect of other polarized sources either in the cluster (radio
sources, radio halo -- $\mu$Gauss magnetic fields and
relativistic particles seem to be ubiquitous in clusters, as
discussed by {\it Wielebinski} --, dust) or along the line of
sight.

\begin{figure}
  \resizebox{\hsize}{25pc}{\includegraphics{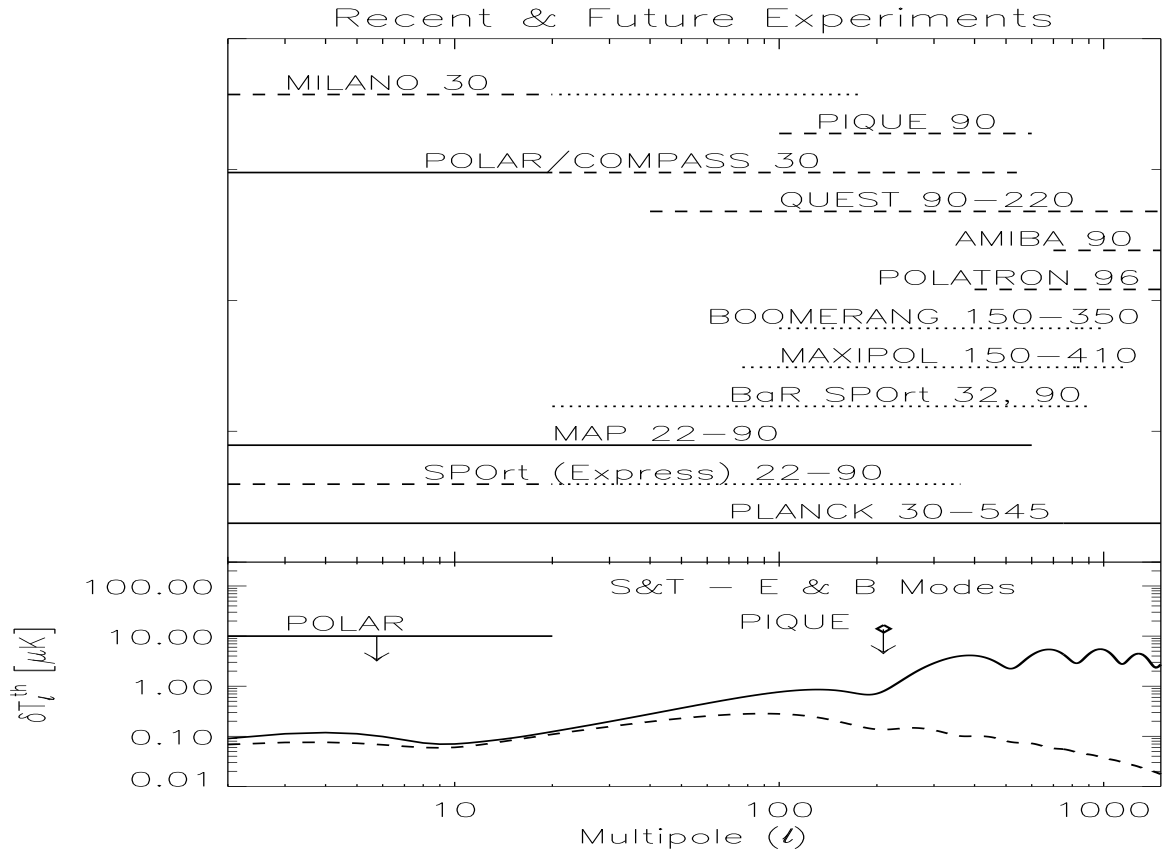}}
\caption{Summary of CMB polarization experiments discussed at this
meeting. The straight lines show the multipole coverage. Note that
the y-coordinate of such lines has nothing to do with the
sensitivity of the instruments. Ground-based experiments are on
top, balloon experiments in the middle and space-borne experiments
on the bottom. The figures close to each experiment give the
observing frequency or the frequency interval covered. The $E$-
(solid) and $B$-mode CMB power spectra (shown for reference) are
for the same model as in Fig.~2. }
\end{figure}

\section{CMB polarization experiments}

An excellent review of ongoing and planned CMB polarization
experiments is given by \cite{Staggs1999}. Figure~4 summarizes
those discussed at this meeting, indicating the multipole range
they cover and the frequency bands they probe.

Although recent experiments (PIQUE \cite{Hedman2001} operating at
$90\,$GHz, and POLAR \cite{Keating2001} at $30\,$GHz) have
considerably improved on previous results, CMB polarization is
still undetected. %PIQUE has however come close to the sensitivity
%required to detect $E$-mode polarization, assuming that its
%amplitude is at the currently expected levels.
As discussed by {\it Timbie}, the upgrade of POLAR, called
COMPASS, which can reach angular scales of $20'$, has the
potential of detecting the CMB polarization if sensitivity per
pixel will be similar to POLAR. CMB polarization measurements are
expected in the next few years from several other experiments. The
NASA Microwave Anisotropy Probe (MAP) is expected to obtain a
highly significant detection of the CMB temperature-polarization
cross-correlation, which is less affected by foregrounds than
polarization maps \cite{KogutHinshaw2000}. Detection of the
$E$-mode polarization is also expected from the new flights of
the balloon-borne experiments, the bolometer-based, now with
polarization capabilities, BOOMERANG ({\it de Bernardis}) and
MAXIMA (MAXIMAPOL), and BAR-SPort ({\it Zannoni, Macculi}), using
radiometers. Very promising ground based experiments include QUEST
({\it Piccirillo}) and  Polatron \cite{Philhour2001}, both using
bolometric receivers, and the interferometer array AMiBA ({\it
Kestevens}, \cite{Lo2001}).

Measuring polarization on large angular scales (cf. the Milano
experiment, presented by {\it Gervasi}, and the SPOrt project, due
to fly on the International Space Station, presented by {\it
Carretti} and {\it Nicastro}) appears to be much more difficult.
The SPOrt-Express project, proposed for reuse of the Mars Express
platform, adds to the SPOrt payload a second instrument, HARI,
equipped with a 50cm antenna, giving an angular resolution $\simeq
0.5^\circ$ at 90 GHz.

{\sc Planck} (see presentations by {\it Delabrouille} and {\it
Villa}) should provide accurate measurements of the $E$-mode power
spectrum up to $\ell \simeq 1500$--2000; its sensitivity would
allow to detect $B$-mode polarization if the ratio of tensor to
scalar contributions to the quadrupole anisotropy is $T/S \gsim
0.045$. Clearly, even if foreground contamination will not prevent
us from reaching these levels, detection of $B$-mode polarization
is by no means guaranteed! Much higher sensitivities are needed to
explore the inflationary parameter space
\cite{KamionkowskiJaffe2000}.

%\subsection{Systematic errors}

As discussed by {\it Kaplan} and {\it Leahy}, systematic errors
may be  most critical. There are a number of potential sources of
systematic errors that could affect the various experiments:
foreground subtraction, beam asymmetries, straylight, electronic
interference, thermal variations, etc.. Experiments measuring
polarization by differencing the outputs from radiometers
sensitive to orthogonal polarization have common systematics for
intensity and polarization measurements, although some
cancellation can be expected in polarization ({\it Leahy}). The
need of combining data from different detectors introduces effects
of calibration, pointing, beam shape mismatches ({\it Kaplan}).
Issues related to the removal of SPOrt systematics have been
discussed by {\it Amisano}. Since the number of physically
interesting parameters is much smaller than the number of
multipoles measured in the power spectrum (typically $\sim 10$
parameters, compared with $\sim 2000$ multipoles in the case of
{\sc Planck}), very small correlated systematic errors which mimic
the $\ell$-dependence of a parameter, can produce a large error in
that parameter \cite{EfstathiouBond1999}.

%\end{itemize}

%\subsection{Sky coverage}

The sky coverage is also an issue. Large area maps are important
to maximize the information from the power spectrum (maps of a
patch of angular size $\theta$ lose the information from modes
with $\ell < 180^\circ/\theta$), to minimize the sample variance
(which scales inversely with the sky coverage), to minimize the
mixing between $B$ and $E$ caused by incomplete sky coverage (due
to the non-local nature of these quantities), which may cause the
$B$ component to be swamped by the, typically much larger, $E$
component \cite{Tegmark2001, Bunn2001}: to detect the $B$
component a survey size much larger than the coherence length of
this component is required. On the other hand, given the extreme
sensitivity required, very long exposures are needed to decrease
the noise per pixel: with realistic integration times only
relatively small patches can be covered. Optimal survey sizes may
be of $\sim 20$--$30^\circ$, with $\sigma_{\rm beam} \simeq
0.3$--$0.5^{\circ}$ \cite{KamionkowskiJaffe2000, Bunn2001,
Lewis2001}.

%\subsection{Data processing}

The flood of data expected from ongoing or forthcoming CMB
experiments is facing us with huge computational challenges. {\it
Balbi} has described an efficient method for constructing
polarization maps. {\it Sbarra} has described a new destriping
technique and discussed its applications with reference to the
SPOrt experiment. The application of a multi-frequency Wiener
filtering technique to polarized maps was investigated by
\cite{Bouchet1999} and \cite{Prunet2000}. The method, generalized
by \cite{Tegmark2000}, requires the knowledge of the average
frequency and angular dependence of the foreground emission. Other
possibilities include pseudo-$C_\ell$ estimates \cite{Wandelt2000}
and harmonic analysis methods that bypass traditional map-based
methods \cite{Challinor2001}. %Clearly, a lot of work is still to
%be done on methods for analysis of CMB polarization data in order
%to derive optimal sky map and power spectrum estimates.

\begin{theacknowledgments}
We are grateful to the organizers of this very timely, useful and
stimulating meeting. Work supported in part by ASI and MIUR.
\end{theacknowledgments}

% choose bibtex style depending on layout style and options used in
% sample:

%\doingARLO[\bibliographystyle{aipproc}]
%          {\ifthenelse{\equal{\AIPcitestyleselect}{num}}
%             {\bibliographystyle{arlonum}}
%             {\bibliographystyle{arlobib}}
%          }
%\bibliographystyle{aipproc}
%\bibliography{Planck}
%\bibliography{sample}

\end{document}